\documentstyle[]{article}
\begin{document}
\begin{center}
{ \huge \bf
On Static Dielectric Response of Microcomposites of the Type: Ferrolastics -
Dielectrics }
\end{center}
\vskip2cm
\begin{center}
{\large  O. Hudak{\footnote{Department of Theoretical Physics and
Didactics of Physics, Faculty of Mathematics, Physics and
Informatics, Comenius University, Bratislava  \\and  \\Department of
Dielectrics, Institute of Physics, CAS, Prague} }  \\
\vskip1cm
Institut fuer Experimentalphysik, University of Vienna, Strudlhofgasse 4, 1090 Wien

\vskip1cm
W. Schranz\\
\vskip1cm
Institut fuer Experimentalphysik, University of Vienna, Strudlhofgasse 4, 1090 Wien.
}
\end{center}

\newpage
\section*{Abstract}
We describe the static dielectric response of ferroelastic-dielectric
microcomposites. Its dependence on temperature, pressure and
concentration is considered for smaller concentrations and for
temperatures above the critical temperature. We have found a
qualitative agreement with existing experimental data for $Al_{2}(WO_{4})_{3}$.

\newpage
\section{Introduction}

The dielectric response of ferroelectric-dielectric type microcomposites
was studied recently \cite{HRP} - \cite{OH2}. When a coupling of the elastic
strain to the electric polarisation is present, then the dielectric
response of such a ferroelastic material may be studied, see in
\cite{jona} - \cite{LG}.  Also the
dielectric response of ferroelastic-dielectric type microcomposites
may be studied. Response of minerals to changing hydrostatic pressure p and
temperature T is also interesting property of these materials. For
example materials of the perovskite type ($LaAlO_{3}$, $CaAlO_{3}$,
$SrAlO_{3}$, $BaTiO_{3}$, $PbNiO_{3}$, $Pb(Zr,Ti)O_{3}$, ...)
undergo a phase transition from a cubic phase to a phase with lower
symmetry at some critical temperature. While a study of
their elastic properties is usually done in literature, we will
consider here the dielectric response of such materials which is due
to coupling between the elastic strain tensor and the polarisation.
 Mechanical analysis is usually done at low
frequencies (0.1Hz - 10 Hz) but also measurements at higher frequencies
are done. Ferroelstic domain wall structure, twinning and other
similar phenomena are studied, see in \cite{STKSS} and \cite{KSSHSS}. Elastic response function (compliance)
shows in Cole-Cole diagrams circular and non-circular behaviour of
these materials in their crystalinne and ceramic form \cite{HRS}. In the
second case multirelaxation phenomena exist in these aterials under
some conditions. Under higher electric and mechanical loading
nonlinear behaviour is exhibited by ferroelectric and ferroelastic
ceramics \cite{E}. While at higher temperatures there is present no
ferroelastic phase, coupling of the elastic strain and electric
polaristion does exist at these temperatures. So dielectricresponse
depends then on mechanical forces acting on the microcomposite of the
ferroelastic-dielectric type. This response enables us to study
properties of these microcomposites using dielectric measurements for
microcomposites under mechanical forces. It is known taht constraint
due to neighbouring material lead below the critical temperature for
transition from paraelastic to ferroelastic phase and due to
shape-change to several forms of the low-temperature phase
\cite{JCD}.  As it is noted by these authors perhaps only in very
small grains there exists a single variant of this form. A strain of
$10^{-4}$ in a grain of size 10 $\mu$m (typical values) is an order of
magnitude too large to be accomodated in a small displacements of
atoms nearby and at surface. An external mechanical field may move
walls between different forms of the low-temperature phase. We assume
in our paper that there are small mechanical fields of the order
$10^{-3}$ and that particles are with heir diameter of the order of
1$\mu$m. They are microcomposites. Moreover in larg concentrations of
the dielectric hard material the clusters of the ferroelastic phase
are still small. Thus we will consider this case in this paper only,
at the percolation transition concentrations there are very large
clusters of ferroelectric and dielectric material, it is not known
today how they are responding to the external mechanical fields. InI
ferroelastic phase long-range anisotropic forces may appear
\cite{LSRSB}. In our paper we discuss properties of the microcomposite
above the critical temperature, thus these long-range forces can be
neglected, they  may be present due to fluctuations of the order
parameter
 but they are of the higher order (so they are small) 

The aim of this paper is to study the static dielectric response of microcomposites of the type: ferrolastics -
dielectrics. In the second chapter a model
of ferroelastic-dielectric 
microcomposites is presented.  We study the dielectric response of
ferroelastic particles.
Ferroelastic particles and their dielectric response to static
electric field in the high
  temperature limit and as a function of hydrostatic pressure is studied in the next chapter. 
Effective Medium Approximation (EMA) is generally formulated for
  dielectric response of microcomposites of the type: ferroelastics -
dielectrics,  and then studied in low concentration limit of the
ferroelastic material and in low concentration limit of the dielectric
material. Microcomposites of the type: ferroelastics -
dielectrics in the percolation region, i.e. for higher concentrations of the
particles but not close to the limiting case $x = 1 $ should be
finally studied however they are from the point of view of mechanical
forces more complicated than the limiting cases and will not be
studied here.
In the last chapter we summarize our results as concerning temperature,
hydrostatic pressure and concentration dependence of the dielectric response of
microcomposites of the type: ferroelastics-dielectrics. A discussion of
these results and their comparison with experiments.

\section{Model of Ferroelastic-Dielectric Microcomposites}

Microcomposites are composited from small
particles. They have interesting properties especially when they consist of two
or more materials with different properties. One of such examples are
ferroelectric-dielectric microcomposites
\cite{HRP} - \cite{OH2}. Ferroelectric properties of particles may appear in
them due
to presence of the polarisation as a primary order parameter. An
interesting possibility is to consider materials in which
ferroelectricity is induced as a secondary order parameter. Primary
order parameter may be the corresponding component of the elastic
strain tensor
(or a combination of components of the elastic strain tensor). The ferroelectric
state is present in such a material due to a coupling between the
elastic strain tensor and the polarisation. Then we have two types of
particles in the microcomposite: ferroelastic and dielectric. Changing
the concentration of these two types of particles in the
microcomposite the response to external fields changes. This holds for
dielectric response and for other type of responses. We are calculating
in this paper the dielectric response of such apredictions
microcomposite. For simplicity we consider all particles in
microcomposite of the same diameter
d. When changing the concentration of the ferroelastic particles
(increasing it)  an infinite ferroelastic cluster of
ferroelastic particles may form. The concentration at which this cluster
forms is the percolation transition concentration. Then increasing
further the concentration the ferroelastic material becomes the matrix
material, and at the concentration $x =1$ one the ferroelastic material only is
formed.

When the concentration of particles changes, the dielectric
response of the microcomposite changes. Thus the interplay between the
percolation transition and the ferroelastic phase
transition appears. In real materials there exists distribution of
diameters of particles and shapes of particles. In our model we neglect this
distribution for simplicity, and consider only spheres.

\section{Ferroelastic Particles: Dielectric Response}

The dielectric response of a ferroelastic particle will be studied
using a
Landau free energy expansion. We will assume cubic symmetry of the
ferroelastic material for simplicity. This symmetry corresponds to
high-temperature symetry of materials mentioned in the Introduction.
For materials of other symmetry
the approach is similar. In ferroelastic materials the primary order
parameter is an elastic strain tensor. Secondary order parameter is 
polarisation.
While the elastic strain tenssor is coupled to the external mechanical fields
(hydrostatic pressure, uniaxial stress, shear stress), the secondary order
parameter is coupled to an external electric field. We will use in our
calculation time and space dependent external fields in general. However for
microcomposites the quasistatic approximation for dielectric response is a convenient
approximation because the wavelength of the electric field is usually much
larger than the diameter of the particle. The space dependence of
the external electric field may be neglected.

To find the free energy expansion we have to find first invariants of the
primary order parameter and of the secondary order parameter, and of
primary and secondary order parameter coupled.
In cubic materials there are the following invariants for the primary
order parameter $\epsilon_{i,j}$, where $i,j = 1,2,3$ denotes axis of
the cubic material (Einstein sum rule is used):

\begin{equation}
\label{1}
\epsilon_{i,j} \epsilon_{j,i} = 3 \epsilon^{2} + 6  \phi^{2}
\end{equation}

which is of the second order in the elastic strain  tensor and:

\begin{equation}
\label{2}
\epsilon_{i,j} \epsilon_{j,k} \epsilon_{k,i} = 3 \epsilon
 (\epsilon^{2} + 2  \phi^{2}) + 6 (2 \epsilon 
\phi + \phi^{2})\phi .
\end{equation}

Here we denoted on-diagonal terms as $ \epsilon $ and off-diagonal
terms of the elastic strain tensor as $ \phi $. 
This term is of the third order in the elastic strain tensor. Because the
coupling between the primary order parameter ( elastic strain tensor ) and the
secondary order parameter - electric polarisation is of the first
order in the elastic strain tensor
and of the
second order in the polarisation, and because of the fourth order in
the electric polarisation order
parameter (or of the sixth order for the first order phase 
transitions near the second order one for the ferroelectric
materials),
we will consider in the free energy expansion only these two invariants 
 for the elastic strain tensor described above, (\ref{1}) and (\ref{2}).
Thus in the corresponding to polarisation part of the free energy
expansion
there will be second order term, fourth order terms and sixth order
terms in general.

The coupling between the elastic strain tensor $\epsilon_{j,k}$ and the
polarisation vector $ P_{i} $ has the form:
 
\begin{equation}
\label{3}
\epsilon_{i,j} P_{i}  P_{j} = \epsilon P^{2}.
\end{equation}

The polarisation vector $P_{i}$ is assumed to have the nonzero component only in the
x-direction, we assume that external electric field will be applied in
this direction.
Then the free energy $F$ expansion has the form:

\begin{equation}
\label{4}
F = \int dV [\frac{B}{2} ( \epsilon^{2} + 2 \phi^{2} ) + \frac{C}{3}(
 \epsilon
 (\epsilon^{2} + 2  \phi^{2}) + 2  (2  \epsilon 
\phi + \phi^{2})\phi ) + \Gamma P^{2}  \epsilon + \frac{\alpha}{2}
 P^{2} + \frac{\beta}{4}
 P^{4} +  \frac{\gamma}{6}
 P^{6} - E.P - \epsilon_{i,j}  \sigma_{j,i}].
\end{equation}

The last term in the free energy expansion depends on the stress
tensor. This tensor may be hydrostatic pressure p, $ \sigma_{j,i}=
\delta_{j,i}p$, uniaxial stress $ \sigma_{x,x}= \sigma $ or shear
stress  $ \sigma_{x,y} $. The constants in the free energy expansion
(\ref{4}) are positive and temperature independent, with the exception
of the constant $B$, for which $ B = B_{0}(T - T_{c})$ where $B_{0}$
is positive, $T_{c}$ is a critical temperature for the transition from
the paraelastic to the ferroelastic phase.

Depending on the field applied (electric, mechanical) we calculate the 
response of the material described by the free energy expansion
(\ref{4}).
We will assume that the surface charge is compensated in the case of
polarized particles. 
The expansion constants in (\ref{4})may be in fact temperature and
hydrostatic pressure dependent. We will not consider this dependence as we
mentioned above with the exception of the constant B. However it may
happen that in some materials these constants are wekly temperature
and hydrostatic pressure dependent. This dependence should be taken into account
when there are  reasons of the experimetal origin.

Let us now consider the high-temperature paraelastic phase.
In this case the free energy expansion from (\ref{4}) takes the
form in which second order terms in the polarisation are taken into
account and the first order terms in the elastic strain tensor are taken into
account. Note that the second order in the electric polarisation is
corresponding to
the first order in the elastic strain tensor.
However the coupling term is of the fourth order in polarisation, so
also the second order term in elastic strain tensor is taken into account.
The fourth order term in polarisation, with the constant beta, is not
taken into account because at high temperatures this term does not
play any essential role.
 Then the free energy $F$ expansion has the form:

\begin{equation}
\label{5}
F = \int dV [\frac{B}{2} ( \epsilon^{2} + 2 \phi^{2} ) +  \Gamma P^{2}  \epsilon + \frac{\alpha}{2}
 P^{2}  - E.P - \epsilon_{i,j}  \sigma_{j,i}].
\end{equation}

As we can see from (\ref{5}) we still may consider different
situations as concerning the influence of the external electric and
mechanical fields.
Let us first consider the static case of the electric field and of the
hydrostatic pressure.

\section{Dielectric Response of Ferroelastic Particles: Static High
  Temperature Limit and Hydrostatic Pressure Dependence}

In this case the free energy  expansion has the form:
 
\begin{equation}
\label{6}
F = \int dV [\frac{B}{2} ( \epsilon^{2} + 2 \phi^{2} ) +  \Gamma P^{2}  \epsilon + \frac{\alpha}{2}
 P^{2}  - E.P - 3 \epsilon p] .
\end{equation}

Here p is the hydrostatic pressure, it is the same in all directions.
The Lagrange-Euler equations for the most stable state at a given
electric field and hydrostatic pressure have the form:

\begin{equation}
\label{6.1}
B \epsilon - 3p + \Gamma P^{2}= 0
\end{equation}
\[ \alpha P + 2 \Gamma \epsilon P - E = 0 \]

Neglecting higher harmonics in the hard dielectrics we obtain that the
dielectric permitivity has the form:

\begin{equation}
\label{6.2}
\epsilon_{f} = \frac{1}{\alpha^{*}} \equiv \frac{1}{\alpha} (1 + \frac{ 6 \Gamma p}{\alpha B})
\end{equation}

As we can see from the equation (\ref{6.2}) the dielectric
permitivity increases with increasing hydrostatic pressure for $ \Gamma $
positive, and it decreases with decreasing hydrostatic pressure for  $ \Gamma $
positive. We assume for simplicity that $ \alpha $ is temperature
independent, and that B is temperature dependent, $ B=B_{0}(T -
T_{0}) \leq 0 $. Thus the dielectric permitivity of the ferroelastic
material is temperature dependent for nonzero hydrostatic pressure. With
increasing temperature  decreases and the dielectric
permitivity tends to the zero hydrostatic pressure value. Note that for
negative $ \Gamma $ there exists a critical hydrostatic pressure $ p_{c} =
\frac{\alpha B}{6 \mid \Gamma \mid} $ at which the permitivity
becomes zero. For hydrostatic pressures higher that the critical the
permitivity is negative. We will not consider the cases in which the
hydrostatic pressure is higher than critical for negative $ \Gamma $.

\section{Effective Medium Approximation - General Formulation for
  Dielectric Response of Composites of the Type: Ferroelastics -
Dielectrics }

The effective medium approximation is valid for the whole interval of
concentrations $ 0 \leq x \leq 1 $. The effective permitivity $
\epsilon_{eff} $  may be obtained from \cite{HRP}:

\begin{equation}
\label{9}
x  \frac{\epsilon_{d} - \epsilon_{eff} }{\epsilon_{d} +2 
  \epsilon_{eff} } + (1-x) \frac{\epsilon_{f} - \epsilon_{eff} }{\epsilon_{f} +2  \epsilon_{eff}}  = 0
\end{equation}

This approximation is based on the response of  a spherical particle to
the whole effective microcomposite. Both components are taken into
account symmetrically. The percolation transition occurs at $ x =
\frac{1}{3} $ for the F component. For the permitivity of the hard
material we take the constant dielectric permitivity $\epsilon_{d}$ 
and $\frac{1}{\alpha}(1 + \frac{ 6 \Gamma p}{\alpha B })$ for the permitivity of the ferroelastic material.
 The effective
permitivity calculated from (\ref{9}) will describe response of the
microcomposite on the electric field and on the hydrostatic pressure. Let us now
consider two limiting case: the limit of small concentration of the
ferroelastic material, and the limit of small concentration of the
hard dielectric material.

\section{Low Concentration of the Ferroelastic Material}

For low concentration of ferroelastic particles in the hard dielectric
material we may calculate the dielectric response of the
microcomposite from (\ref{9}). In this case  
 a small parameter is $ 1 - x $. The effective dielectric
permitivity $ \epsilon_{eff}$ has the form:

\begin{equation}
\label{10}
\epsilon_{eff} = \epsilon_{d} + 3(1 - x) \epsilon_{d}
\frac{\frac{1}{\alpha^{*}} - \epsilon_{d}}{\frac{1}{\alpha^{*}} +2 \epsilon_{d} }
\end{equation}

Substituting for the ferroelastic permitivity from (\ref{6.2})we
obtain hydrostatic pressure, temperature and concentration dependence of the
effective dielectric permitivity in the limit $ x$ near the value 1:

\begin{equation}
\label{11}
\epsilon_{eff} = \epsilon_{d} + 3(1 - x) \epsilon_{d}. 
\end{equation}
\[ . \frac{\frac{1}{\alpha}(1 + \frac{ 6 \Gamma p}{\alpha B} ) -
  \epsilon_{d}}{\frac{1}{\alpha}(1 + \frac{ 6 \Gamma p}{\alpha B} ) + 2 \epsilon_{d} } \]

As we can see with increasing concentration of the ferroelastic
component, decreasing x, the dielectric response increases
assuming that the permitivity of the hard material is lower than that
of the ferroelastic material. 
If the permitivity of the hard material is higher than that
of the ferroelastic material, then with increasing concentration of
the ferroelastic component the effective dielectric permitivity decreases.
Increasing temperature the effective dielectric
response of the microcomposite is tending to that of the dielectric
response of microcomposite with zero  hydrostatic pressure. Increasing
hydrostatic pressure to higher values the effective dielectric response becomes
less hydrostatic pressure dependent.

\section{Low
  Concentration of the Dielectric Material}

For low concentration of dielectric particles in the ferroelastic matrix
 we may calculate the dielectric response of the
microcomposite again from (\ref{9}) in a similar way as in the
preceeding section. In this case we have $ x $ near the value 0.
Thus a small parameter is now $ x $. The effective dielectric
permitivity $ \epsilon_{eff}$ is in this case given as:

\begin{equation}
\label{12}
\epsilon_{eff} = \epsilon_{f} + 3 x \epsilon_{f} 
\frac{\epsilon_{d} -  \epsilon_{f}}{\epsilon_{d} + 2 \epsilon_{f} }
\end{equation}

Substituting for the ferroelastic permitivity from (\ref{6.2})we
obtain hydrostatic pressure, temperature and concentration dependence of the
effective dielectric permitivity in this limit as:

\begin{equation}
\label{13}
\epsilon_{eff} = \frac{1}{\alpha}(1 + \frac{ 6 \Gamma p}{\alpha B }) + 
\end{equation}
\[ + 3 x \frac{1}{\alpha}(1 + \frac{ 6 \Gamma p}{\alpha B }).
\frac{\epsilon_{d} - \frac{1}{\alpha}(1 + \frac{ 6 \Gamma p}{\alpha B})}{
\epsilon_{d} + 2 \frac{1}{\alpha}(1 + \frac{ 6 \Gamma p}{\alpha B })} \]

\section{Summary}

We studied here the static dielectric response of microcomposites:
ferroelastics-dielectrcs. A model for such a microcomposite was
formulated. Dielectric properties of ferroelastic particles were
studied. Cubic symmetry of the material was assumed. This corresponds
to most of the known ferroelastic materials at high temperatures.
Materials with other symmetry may be studied in a similar way. We
consider a coupling of the elastic strain tensor to the electric
polarization.
While the primary order parameter is coupled to external mechanical
fields, the secondary order parameter is coupled to external electric
field. For microcomposites a quasistatic approximation may be used for
the dielectric response. We have found the free energy expansion using
invariants of the primary order parameter, of the secondary order
parameter, and of their mixed terms. We assumed that in the free
energy expansion only the coefficient B of the second order of the
primary order parameter is temperature dependent, and that other
parameters are temperature independent. All of these coefficients are
assumed to be hydrostatic pressure independent. Face charges are assumed to
compensate the dipole moment in ferroelastic particles in which an
electric dipol appears. We apply hydrostatic pressure on
the microcomposite and consider high-temperature properties of such a
microcomposite here. The low-temperature properties may be considered
in the same way easily. The most important role in the free
energy expansion play terms of the second order in the elastic strain tensor
and of the second order in electric polarisation. The fourth order
terms of the polarisation are small and we do not consider them.
Firstly we studied the dielectric response of the ferroelastic particles
for high temperatures (temparatures above the transition temperature
to the ferroelstic phase) and the hydrostatic pressure dependence.
We have found the dielectric constant in the static limit here, which
correspond to mechanical measurements from 0.1Hz to 10 Hz. For higher
frequencies the dynamic dielectric constant may be calculated
directly.
It turns out tha the dielectric response of a particle is depending on hydrostatic pressure:
incresing hydrostatic pressure the response is larger (for the constant $\Gamma$
positive, which is the most usual case), its increase is linear in
hydrostatic pressure.
The nearer we are to the ferroelastic transition temperature the
higher is the change in the dielectric response for the same change of
hydrostatic pressure.
At the transition temperature the dielectric response constant
diverges in this static limit for nozero hydrostatic pressures. The dielectric
response of a particle decreases increasing hydrostatic pressure for the constant
$ \Gamma $ negative. We consider only such hydrostatic pressures in this case
which are above the critical hydrostatic pressure $ p_{c} = \frac{\alpha B}{6 \mid
  \Gamma \mid} $. The effective medium approximation theory is
formulated for dielectric response of the microcomposite. The two
limiting cases are considered: the limit of small concentration of the
ferroelastic particles, and the limit of small concentrations of the
hard dielectric material. The effective dielectric constant is
calculated in both cases.
In the first case we have found that
increasing the concentration of the ferroelastic component the
dielectric response increases assuming that the permitivity of the
hard material is lower than that of the ferroelastic. In the oposite
case it decreases. Incresing temperature the effective dielectric
constant of the microcomposite is tending to the dielectric response
of the microcomposite which is under no hydrostatic pressure. If
temperature decreses to the transition temperature from paraelastic to
ferroelastic phase then the effective response is becoming independent
on the
hydrostatic pressure. For temperatures higher that the critical it is
hydrostatic pressure dependent. For small hydrostatic pressures
linearly, for larger hydrostatic pressures
and positive $\Gamma$ the effective response increases, however with
smaller and smaller velocity. 
In the second case of low concentration of the dielectric material the
effective permitivity increases with hydrostatic pressure for positive constant $
\Gamma $ . Increasing the concentration of the dielectric hard
material the effective dielectric constant decreases if the
ferroelastic dielectric constant is higher than that of the dielectric
constant. For high temperatures the effective dielectric constant
becomes such as if no hydrostatic pressure was applied. For temperature decreasing
to critical temperature the effective ferroelectric constant becomes
very sensitive to hydrostatic pressure values however with larger concentrations
of the dielectric hard material this sensitivity is lower.

We described a theory of microcomposites of the ferroelastic-dielectric
type in the limiting cases of concentrations. Quasistatic
approximation is used.
  The Effective Medium Approximation as described
above is less usefull for description of the region of percolation
transition. For the composite theory of the elastics-elastics composite, which is
of the EMA type, see in \cite{B}.
In $Sn-VO_{2}$ composites
for example \cite{JL} 5 percents of inclusions displayed various
mechanical instabilities, and 0.5 volume percents of inclusion shows
no signs of such instabilities. According to authors this
concentration dependence is in agreement with composite theory, which is
of the EMA type, see in \cite{B}.

In our model above we did not consider the mechanical
inclusion/matrix interactions. The ferroelastic crystals
 as ferroelastic particles may have domain wall structure. However we
 discuss temperatures in which there is no transition to the
 ferroelastic phase, the paraelastic phase. Effect of elastic clamping
 was not conisdered here. For its discussion for an improper and
 a pseudoproper ferroelastic inclusion see \cite{PS}. For
 elipsoidal shapes of the ferroelastic inclusions the order parameter and
 strain are uniform inside the inclusion. For improper and
 pseudoproper ferroelastic inclusions and polycrystals
 inclusion/matrix interaction renormalizes the constant of the Landau
 free energy expansion of the order parameter. For proper ferroelastic
 materials which we consider here 3D clamping of the crystal inclusion
 in the matrix is not considered, we assume the mechanical
 equilibrium of the inclusion/matrix system is present at temperatures
 higher than the critical temperature.
Results of measurements of low frequency dielectric constant and dielectric loss on
orthorhombic $ Al_{2}(WO_{4})_{3}$ show \cite{MVATG} for polycrystalline material,
where voids may play the role of dielectric material, linear
increasing dependence with hydrostatic pressure. This is qualitatively in
agreemnet with our theory. In epitaxial films proper ferroelastic
phase transition with symmetry-conserving and symmetry-breaking misfit
strains may be present \cite{BL1}. Most of microcomposites are in the form of
films. The authors \cite{BL1} have found that if the extrinsic misfit
strain,   does not break the symmetry of the high-temperature phase,
the transition in the film
occurs at somewhat lower temperature than in the bulk. This phenomenon
we neglect in our paper. In cubic-tetragonal systems like $Nb_{3}Sn$,
$V_{3}Si$, $In-Tl$ alloys, $Fe-Pd$ alloys and $ Ni_{2}MnGa$ the cubic
cell elongates (or contracts) in one of the main axis to form tetragonal
cell \cite{CJ}. While these interesting materials are in general
different from those about which we wrote in the introduction above,
their high-temperature phase is cubic and our approach may be used to
study them.  Results of
our paper may be interesting for interpretation of experiments on
ferroelastic-dielectric microcomposites.

\section*{Acknowledgement}

One of the authors (O.H.) wishes to express his sincere thanks to V.Dvorak, J.Petzelt,
I.Rychetsky, J.Holakovsky and M.Glogarova from the Institut of
Physics, CAS, Prague for their
discussions on microcomposite materials. He also thanks to
Prof. A. Fuith and Prof. W. Schranz 
for their kind hospitality during his stay in Vienna. 
This paper was supported by the grant  VEGA-project 1/0250/03 and
OEAD-WTZ Austria-Czech Republic cooperation project No. 2004/24.

\end{document}